\documentclass{PoS}

\title{Event-by-event weighting at next-to-leading order}

\ShortTitle{Event-by-event weighting at NLO}

\author{\speaker{Ciaran Williams}\\
	 \thanks{Preprint number: FERMILAB-CONF-13-531-T }
    Niels Bohr International Academy and Discovery Center,   \\ University of Copenhagen, Blegdamsvej 17, DK-2100 N, Copenhagen, Denmark  \\
    E-mail: \email{ciaran@nbi.dk}}

\author{{John M. Campbell, Walter T. Giele}\\

        Fermilab, Batavia IL, 60510 USA\\
        E-mail: \email{johnmc@fnal.gov, giele@fnal.gov}}

\abstract{
We present a general method of associating next-to-leading order weights to leading order phase space configurations
at hadron colliders. The method relies on a re-organization of 
phase space for the real radiation contributions, defining a one-to-many map such that 
each point in the real phase space is associated with a distinct Born topology. As a result virtual and 
real singularities cancel at each Born phase space point. The new phase space maps can be used in a traditional 
approach for NLO calculations. However, interesting applications arise when one instead integrates out the real radiation 
up to a specified scale. This allows one to define NLO weights for individual phase space points that are present
at LO. This method allows for the extension of matrix element methods to next-to-leading
order, even in the presence of jets. We discuss some recent applications of the matrix element method 
to Higgs physics at the LHC.
}

\FullConference{11th International Symposium on Radiative Corrections (Applications of Quantum Field Theory to Phenomenology) (RADCOR 2013),\\
		22-27 September 2013\\
		Lumley Castle Hotel, Durham, UK }

\begin{document}

\section{Introduction}

In order to search for exotic new physics, or to accurately measure the 
properties of known particles, one must extract the maximal amount of physical information from each event.
One way of achieving this goal is to assign each event a weight associated 
with a chosen theoretical hypothesis. These event by event discriminants can offer 
improvements over more traditional analyses that are based on a sequence of kinematic cuts.
The Matrix Element Method (MEM)~\cite{Kondo:1991dw,Dalitz:1991wa,Fiedler:2010sg,Artoisenet:2010cn,Alwall:2010cq}
represents a class of kinematic discriminants in which a fixed order matrix element is used 
to assign the theoretical weight. This is motivated by the fact that, by its very nature, the matrix element contains the most
theoretical information available. Historically, a drawback of the MEM was its restriction to leading order
matrix elements. Extensions beyond LO are necessary  in order to ensure the theoretical rigor of the method,
but require careful consideration of the real and virtual phase spaces and their associated singularities.

The aim of this work is to provide a mechanism of combining real and virtual phase spaces such that event-by-event 
weights can be defined at NLO. The resulting algorithms can then be used to construct a MEM accurate to NLO.
We will define a method that provides unique NLO weights for individual Born phase space points.
Section~\ref{sec:EVE@NLO} describes how the weights are constructed, beginning from a LO topology. 
We present some simple validations of the method in section~\ref{sec:Val}, and discuss the application to the 
MEM in section~\ref{sec:MEM@NLO}. Finally in section~\ref{sec:Conc} we present our conclusions.

\section{Event-by-event weighting at NLO}
\label{sec:EVE@NLO}

\subsection{Electroweak final states} 
 
We begin by presenting the simplest
case in which the LO event contains no colored final state particles. These electroweak final states were studied in detail 
in the context in ref.~\cite{Campbell:2012cz}, which presented a new implementation of the MEM at NLO using event-by-event
NLO reweighting. In order to produce well-defined and unique weights
the MEM requires integration over the initial state longitudinal degrees of freedom. This integration is a feature of the MEM method and not a core 
requirement of the event-by-event reweighting. Therefore in the following section we simplify the discussion by focussing solely on the reweighting 
of LO phase space points to NLO. The aim of this section is to define an event-by-event $K$-factor such that the NLO calculation is rendered in the following format, 
\begin{eqnarray}
\mathcal{P}_{NLO}(\Phi_B) = K(\Phi_B) \mathcal{P}_{LO}(\Phi_B).
\end{eqnarray}
Here $\mathcal{P}(\Phi_B)$ represents a weight defined at a given order for an input Born phase space point,$\Phi_B$.  
The Born phase space point is defined as follows, 
$\Phi_{B}=(x_1,x_2,\{Q_n\})$,
where $\{Q_n\}$ is a set of four momenta which represent the $n$ final state electroweak particles. The two initial partons 
are defined in the lab frame and are fully specified by the fractions of the beam momentum, $x_1$ and $x_2$.
Given this phase space point it is trivial to define a weight using the LO matrix element. 
\begin{eqnarray}
\mathcal{P}_{LO}(\Phi_B)= \frac{f(x_1)f(x_2)}{2x_1x_2s}|\mathcal{M}^{(0)}(\Phi_B)|^2
\label{eq:LOWT}
\end{eqnarray}
Upon integration over the full Born phase space one reproduces the LO cross section, i.e. 
\begin{eqnarray}
\sigma_{LO}=\int d x_1 \, d x_2  \prod_{i=1}^{n} d^4 p_i \;\delta^{(+)}(p_i^2-m_i^2) \;\delta^{(4)}(\sum_{i} p_i - p_1 -p_2)   \;\mathcal{P}_{LO}(\Phi_B) 
\end{eqnarray}
We now wish to define the NLO corrections to this fully exclusive phase space point $\Phi_B$. This is trivial to evaluate for the virtual corrections, since they
share the same phase space
\begin{eqnarray}
\tilde{\mathcal{{P}}}_{V}(\Phi_B)=\frac{f(x_1)f(x_2)}{2x_1x_2s}\bigg(|\mathcal{M}^{(0)}(\Phi_B)|^2+2\rm{Re}\bigg\{\mathcal{M}^{(0)}\mathcal{M}^{(1)^{\dagger}}(\Phi_B)\bigg\}\bigg)
\end{eqnarray}
In this equation we introduced the notation $\tilde{\mathcal{{P}}}_{V}$, where the tilde signifies a weight that is divergent.
Weights without a tilde have been rendered finite, using a prescription that we will describe shortly.  

Having defined our weights for the virtual corrections our remaining task is to evaluate the real corrections which occupy  the larger phase space $\Phi_R$. 

Our aim is to define a map between the real and Born phase spaces such that the two calculations can be combined together in a meaningful way.
We thus choose to define the real phase space as a map from the Born kinematics using the following definition  
$\Phi_R(\Phi_B) = (x_a,x_b,\{Q_n\},p_r)$
where $Q_n$ corresponds to the electroweak particles associated with the original Born point.  In this formalism it is clear that all final state (electroweak) Lorentz
invariant quantities are preserved between the Born and the real phase space points.  It is also clear that it is impossible to maintain $Q_n$ and momentum conservation
whilst maintaining collisions along the $z$ axis. Our setup requires the former, so it is necessary to move the initial state away from the $z$-axis. For this reason, it is clear that
quantities that are not Lorentz invariant take different values in the two phase spaces.
The lab frame is restored by boosting the new phase space point back to a frame in which the beams are longitudinal, and it is in this frame in which the parton fractions are defined and PDFs are evaluated.  
Fully inclusive NLO cross sections are obtained by integrating the emissions over the full phase space.
Exclusive NLO cross sections are defined by integrating up to a $p_T$ scale in which the parton would be observed as a (lab frame) jet.   

In this formalism it is natural to use a forward branching phase space generator~\cite{Giele:2011tm} in which a Born phase space point undergoes
a branching in the initial state to produce the real radiation. The corresponding element of phase space is,  
\begin{eqnarray}
d\,\Phi^{IS}_{\mbox{\tiny FBPS}}=\frac{1}{(2\pi)^3}\frac{Q^2}{s_{ab}} d\,t_{ar} d\,t_{rb} d\,\phi
\end{eqnarray}
where $a$ and $b$ represent the new initial state momenta, and $p_r$ is the branched momenta. 
Using this phase space we can define the following real weight 
\begin{eqnarray}
\tilde{\mathcal{P}}_{R}(\Phi_B)=\int d\,\Phi^{IS}_{\mbox{\tiny FBPS}}(\Phi_B) J_x \frac{f(x_a)f(x_b)}{2x_ax_b s} |M^{(0)}_R(\Phi_R(\Phi_B))|^2
\end{eqnarray}
where $J_x$ represents the Jacobian from changing the initial state variables $(x_1,x_2)$ to $(x_a,x_b)$. In our setup, where we integrate
over $x_1$ this factor is given by $J_x=1/(x_1s)$. Note that, at this stage, our weight is still divergent. 

\subsubsection{Regulating the cross section} 

Since $\tilde{\mathcal{P}}_R$ and $\tilde{\mathcal{P}}_V$ are separately divergent they need to be regulated in order to define our physical, dynamical $K$ factors.
In order to do this, our regulating procedure for the real radiation must only involve the phase space point at which the virtual contribution
is evaluated.  This rules out the possibility of using the Catani-Seymour dipole procedure~\cite{Catani:1996vz}, in which singularities are
cancelled by mapping an individual real phase space point to many Born configurations. 
However a simple alternative is to use phase space slicing~\cite{Giele:1991vf,Giele:1993dj}. This method introduces a small parameter $s_{min}$ and
one integrates the full real matrix element under the condition that all $s_{ij}$ lie above this threshold.
Below this threshold one integrates simplified functions which reproduce the soft and collinear singularities of the full matrix element.
This procedure is accurate to $\mathcal{O}(s_{min})$, so is a good approximation for sufficiently small $s_{min}$.
The simplified matrix elements are simple enough to integrate analytically, producing counter terms which cancel the remaining poles in the virtual amplitude. 
 this regulating prescription we can define the weight as follows, 
\begin{eqnarray}
\mathcal{P}_{NLO}=\frac{f(x_1)f(x_2)}{2x_1x_2s}\bigg((1+\mathcal{R}_v(s_{min}))|\mathcal{M}^{(0)}(\Phi_B)|^2+2\rm{Re}\bigg\{\mathcal{M}^{(0)}\mathcal{M}^{(1)^{\dagger}}(\Phi_B)\bigg\}\bigg)\nonumber\\+\int_{s_{min}} d\,\Phi^{IS}_{\mbox{\tiny FBPS}}(\Phi_B) J_x \frac{f(x_a)f(x_b)}{2x_ax_b s} |M^{(0)}_R(\Phi_R(\Phi_B))|^2+\mathcal{O}(s_{min})
\label{eq:NLOWT}
\end{eqnarray}
In this equation $\mathcal{R}_v$ corresponds to the integrated real phase space and, since real radiation is subject to the cutoff parameter, it depends on $s_{min}$.

\subsection{Final states with jets} 

We now wish to extend the results of the previous section to include Born final states which contain jets. We define a
jet using the following kinematic variables, 
\begin{eqnarray}
J_i=(p_{T,i},\eta_i,\phi_i,m_i) 
\end{eqnarray} 
where $p_T$,  $\eta$,  $\phi$ and $m_j$ represent, respectively, the transverse momentum, pseudo-rapidity, azimuthal angle and mass of the jet.
At Born level a jet has a mass equal to its parent parton, which in here we take to be zero. We can now define a Born phase space point as follows,
$\Phi_B=(x_1,x_2,\{Q_i\},\{J_j\})$
where the $n$-particle final state is determined by $i$ electroweak particles ($\{Q_i\}$) and $j$ jets ($\{J_j\}$). In order to assign a fixed order weight to this phase space point
we must define the function which maps the four vectors of $m$ partons to $n$ jets. This jet-function $C(\{p_m\},\{J_n\})$ is crucial in order to define exclusive weights for events containing jets. At LO the jet-function is particularly simple since each parton can be assigned to an individual jet 
\begin{eqnarray}
C^{LO}_{1\vert m}(\{p_{m}\}|\{J_{m}\})=\prod_{i=1}^m \delta(p_{T,i}-p^J_{T,i})\delta(\phi_i-\phi^J_i) \delta(\eta_i-\eta^J_i) \end{eqnarray}
We note that the product of jet functions are themselves jet functions, 
\begin{eqnarray}
C^{LO}(\{p_1 \ldots p_i\}|\{J_1 \ldots J_i\})C^{LO}(\{p_{i+1} \ldots p_m\}|\{J_{i+1} \ldots J_m\})=C^{LO}(\{p_{m}\}|\{J_{m}\})
\end{eqnarray}
Using our jet function we can define our LO weight in the presence of jets 
\begin{eqnarray}
\mathcal{P}_{LO}(\Phi_B)= \frac{f(x_1)f(x_2)}{2x_1x_2s}|\mathcal{M}^{(0)}(\Phi_B)|^2C^{LO}(\{p_{m\}}|\{J_{m}\}).
\label{eq:LOWTjet}
\end{eqnarray}
As was the case in the previous section it is trivial to include the virtual corrections at this phase space point, since they share both the phase space and jet-functions. We will also proceed to regulate our virtual amplitudes using the integrated approximate matrix element from the slicing setup. Our virtual weight is thus defined as 
\begin{eqnarray}
\mathcal{P}_V=\frac{f(x_1)f(x_2)}{2x_1x_2s}\bigg((1+\mathcal{R}_v(s_{min}))|\mathcal{M}^{(0)}(\Phi_B)|^2+2{\rm{Re}}\bigg\{\mathcal{M}^{(0)}\mathcal{M}^{(1)^{\dagger}}(\Phi_B)\bigg\}\bigg)C^{LO}(\{p_{m}\}|\{J_{m}\})\nonumber\\
\label{eq:VWTjet}
\end{eqnarray}
The jet-function for the real radiation maps an $m+1$ parton level event to $m$ jets, in  one of two ways.
Firstly the jets could be identified with individual partons, in the same manner as at LO. Secondly, a new topology for the real phase space occurs when two partons are clustered together to form a jet. The real jet-function is thus defined as follows, 
\begin{eqnarray}
C(\{p_1\ldots p_{m+1}\}|\{J_1\dots J_{m}\})&=&\sum_{i=1}^{m+1} C^{LO}(\{p_1 \ldots p_{i-1}\}|\{J_1 \ldots J_{i-1}\})\;
 C^{LO}(\{p_{i+1} \ldots p_{m+1}\}|\{J_{i+1} \ldots J_{m+1}\})\nonumber\\&+&\sum_{i=1}^{m}\sum_{j=i+1}^{m+1} 
\left(\prod_{\alpha=1,\alpha \ne i,j}^{m+1} \delta(p_{T,\alpha}-p^J_{T,\alpha})\delta(\phi_\alpha-\phi^J_\alpha) \delta(\eta_\alpha-\eta^J_\alpha)\right)\nonumber\\&&\times
 \delta(p_{T,i+j}-p^J_{T,i})\delta(\phi_{i+j}-\phi^J_i) \delta(\eta_{i+j}-\eta^J_i)\nonumber\\
&=&\sum_{i=1}^{m+1} C_{IS}(i) + \sum_{i=1}^{m}\sum_{j=i+1}^{m+1} C_{FS}(i,j) 
\end{eqnarray}
The first term in the above equation, in which individual partons are identified as final state jets, is clearly very similar to the situation described in the previous
section. We define this region as $C_{IS}(i)$, where the $i$ denotes the parton which is not associated with a jet. It is most natural in this region to use the initial
state forward brancher, with the branched parton being identified as particle $i$. The second summation, over $C_{FS}(i,j)$ is a new feature at NLO, and represents 
configurations in which two partons $p_i$ and $p_j$ cluster to form the jet $J_{\alpha}$. The clustering of the two partons results in the jet acquiring a non-zero
mass. Since our jet definition freezes $p_T$, $\eta$ and $\phi$ it is clear that the NLO jet is related to the LO jet by a longitudinal rescaling.

Using this setup it is straightforward to construct a final state forward branching phase space generator.
This phase space generator branches a Born jet to produce two massless partons.  Provided they pass the clustering algorithm, they automatically
cluster to reproduce the kinematic properties of the Born jet, whilst having a non-zero mass. Specifically the phase space factorizes as follows, 
\begin{eqnarray}
d\Phi_R(x_a,x_b,\{Q_n\},\{p_{m+1}\}) \rightarrow \sum_{ij,\alpha}d\Phi_B(x_1,x_2,\{Q_n\},\{J_{m-1}\}) d\,\Phi^{FS}_{\mbox{\tiny FBPS}}(p_i,p_j,J_{\alpha})
\end{eqnarray}
The momenta $p_i$ and $p_j$ are generated according to the final state forward branching measure, 
\begin{eqnarray}
d\,\Phi^{FS}_{\mbox{\tiny FBPS}}(p_i,p_j,J_{\alpha})=\frac{m_Tm_L}{J_{L}\cdot p_{jL}} d p_i^{(4)}  \delta^{(+)}(p_i)^2
\end{eqnarray}
Given the vector $p_i$, $p_j$ is constrained such that $p_i+p_j$ has the same values of $p_T$, $\eta$ and $\phi$ as the Born jet $J_{\alpha}$. In order to ensure
momentum conservation the beam undergoes a longitudinal Lorentz transformation.  The quantities $m_T$ and $m_L$ are the transverse and longitudinal mass of
$p_i+p_j$. If partons $i$ and $j$ are not selected to cluster under the jet algorithm then the event is rejected. 

We can combine the results of this section to define a NLO weight
for an exclusive final state including jets, 
\begin{eqnarray}
\mathcal{P}_{NLO}=\frac{f(x_1)f(x_2)}{2x_1x_2s}\bigg((1+\mathcal{R}_v(s_{min}))|\mathcal{M}^{(0)}(\Phi_B)|^2+2\rm{Re}\bigg\{\mathcal{M}^{(0)}\mathcal{M}^{(1)^{\dagger}}(\Phi_B)\bigg\}\bigg)\nonumber\\+\sum_{i=1}^{n_{jets+1}}\int_{s_{min}} d\,\Phi^{IS}_{\mbox{\tiny FBPS}}(\Phi_B) J_x \frac{f(x_a)f(x_b)}{2x_ax_b s} |M^{(0)}_R(\Phi_R(\Phi_B))|^2C_{IS}(i) \nonumber\\+\sum_{i=1}^{n_{jets}}\sum_{j=i+1}^{n_{jets}+1}\int_{s_{min}} d\,\Phi^{FS}_{\mbox{\tiny FBPS}}(\Phi_B) J_x \frac{f(x_a)f(x_b)}{2x_ax_b s} |M^{(0)}_R(\Phi_R(\Phi_B))|^2C_{FS}(i,j)+\mathcal{O}(s_{min})
\label{eq:NLOWTjet}
\end{eqnarray}
A dynamical $K$-factor is then given by the ratio $\mathcal{P}_{NLO}/\mathcal{P}_{LO}$, which
is defined point-by-point in the Born phase space.

\section{Validation}
\label{sec:Val}
\begin{figure}
\begin{center}
\includegraphics[width=6.2cm]{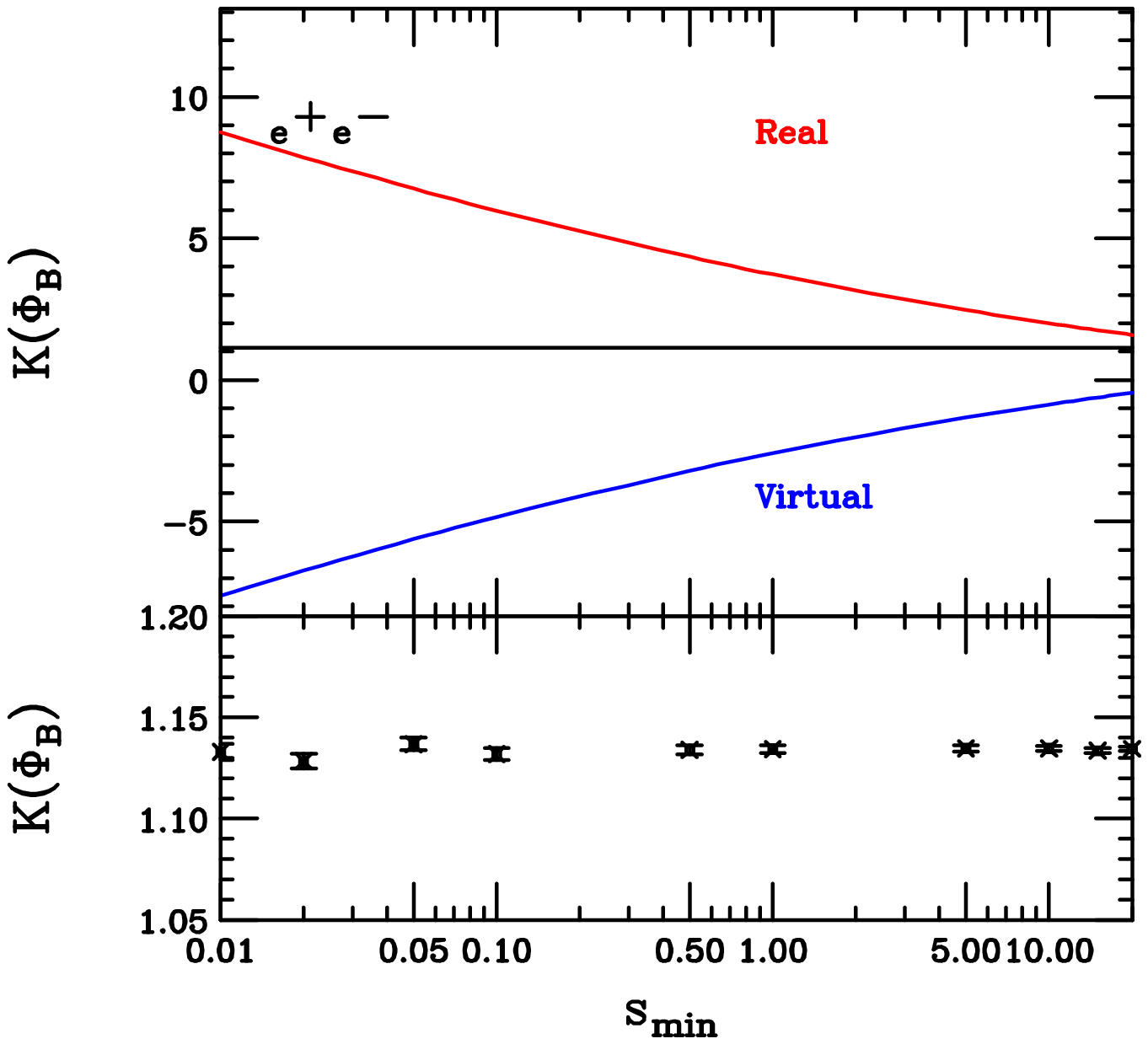}
\includegraphics[width=6.2cm]{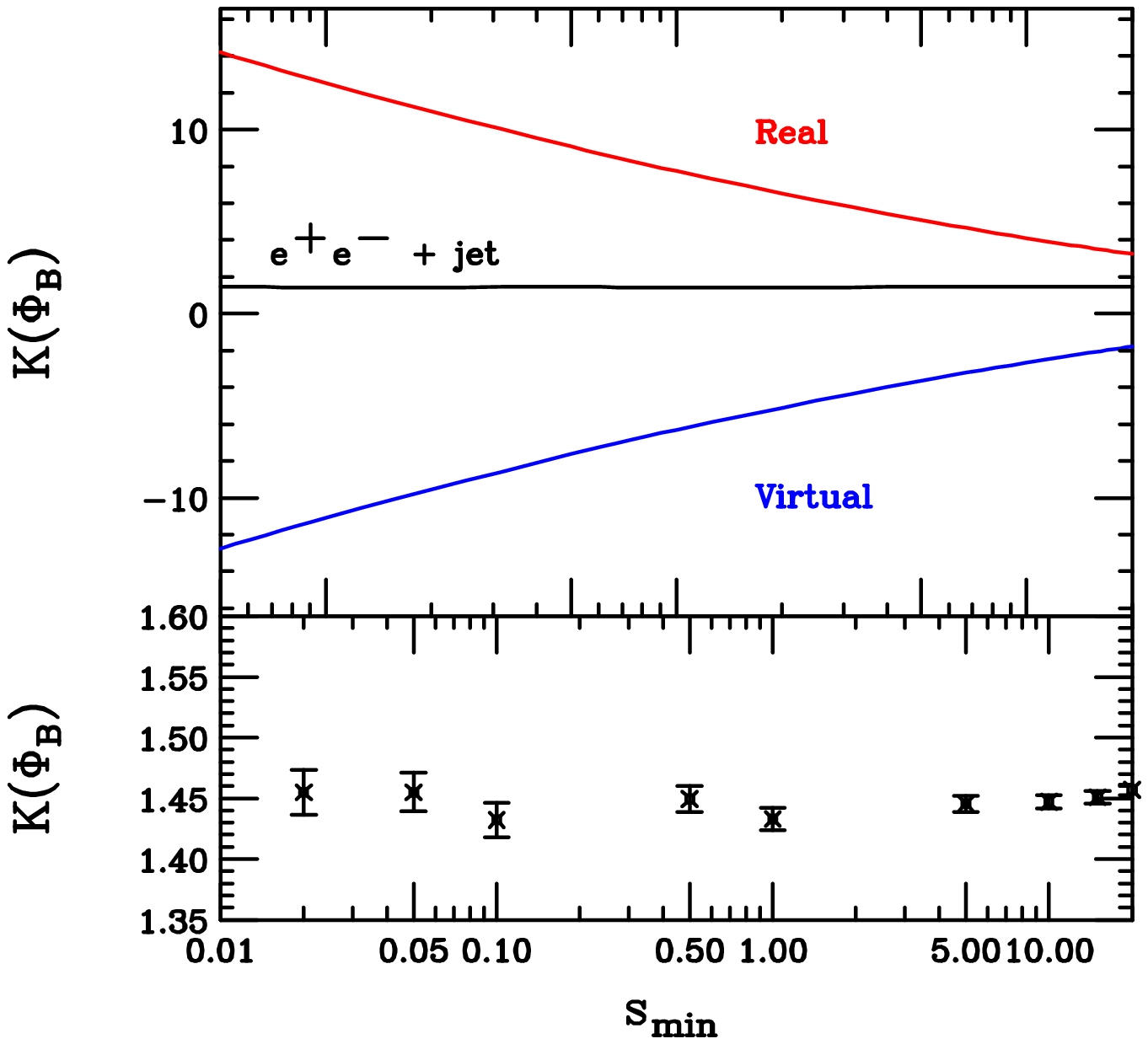}
\caption{The dependence of the dynamical $K$ factor on $s_{min}$, for $pp \to e^+e^-$ (left) and $pp \to e^+e^- + $jet (right).
In each calculation the $K$ factor is calculated for a single phase space point.}
\label{fig:Zsmin}
\end{center}
\end{figure}

We validate our method by investigating the dependence of the NLO cross section on the regulating parameter $s_{min}$,
at a fixed Born phase space point.  We consider both the simplest case of an electroweak final state, $pp \to e^+e^-$,
as well as the extension to include a jet, $pp \to e^+e^- + $jet. 
Our results, for sample phase space points, are shown in Fig.~\ref{fig:Zsmin}.  The 
cancellation of logarithms between the virtual and real contributions is apparent in both cases, resulting in 
a dynamical $K$ factor that shows no significant dependence on $s_{min}$.
We have also checked that, if the full differential information regarding $\mathcal{P}^{NLO}$ is maintained,
we reproduce the results of a traditional NLO calculation when integrating over the entire Born phase space.
\section{The Matrix Element Method at NLO}
\label{sec:MEM@NLO}

An obvious application of the method described above is the extension of the MEM to NLO accuracy. In this 
section we provide a brief introduction to this application. The MEM proceeds by associating an input event with a weight arising 
from the matrix element.  For simplicity, consider the production of a $Z$ boson in a hadronic collision.
Since such events in data, or those obtained from Monte Carlo event simulations, typically possess significant hadronic recoil,
one must map this onto a LO partonic configuration in which transverse momentum is balanced.
An appropriate map is obtained by boosting the event such that the transverse momentum of the LO final state balances. 
Since there are multiple longitudinally-connected Lorentz transformations that result in the same transverse 
final state, this boost is not unique.  One must therefore integrate over all longitudinally equivalent, boosts.
This is accomplished by integrating over the parton fractions $x_i$, 
under the constraint that the invariant mass of the final state is preserved. The MEM weight is then defined by,
\begin{eqnarray}
P^{MEM}(\tilde{\phi}_B)=\frac{1}{\sigma}\int d x_1 d\phi_B \, \mathcal{P}(\phi_B) W(\phi_B,\tilde{\phi}_B)
\end{eqnarray}
In the above equation the input event ($\tilde{\phi}_B$) is related to a parton level event $(\phi_B)$ by the transfer functions $W$. 
The Lorentz transformation is then applied to the parton level event $\phi_B$, and the longitudinal integration occurs for each generated
phase space point. 
The detailed discussion of the transfer functions, $W$ is beyond the scope of this work. Here we merely note that they are rather simple
for well-identified final state leptons, and much more complicated for final state jets. 
Using the results of the section~\ref{sec:EVE@NLO} it is fairly straightforward to extend the MEM
to NLO, by simply using the appropriately-defined $\mathcal{P}$ and $\sigma$.
It is also possible for the transfer functions $W$ to change when moving from LO to NLO. Such modifications can be estimated 
in data by comparing the use of parton showers that include both LO and NLO effects to fit the transfer functions. Ideally the latter
option would be used for a consistent picture at NLO.
Using these weights one can construct event-by-event kinematic discriminants to search for events corresponding to particular choices 
of the matrix element, for instance corresponding to signal and background processes.


In ref.~\cite{Campbell:2013hz} the MEM@LO and MEM@NLO 
were used to study the decay of a Higgs boson into the $Z\gamma$ final state. This 
rare decay of the Higgs boson is incredibly difficult to search for experimentally due to the combination 
of low rate and kinematic similarities between the signal and backgrounds. 
This search is a therefore a prime candidate to benefit from the additional power of MEM discriminants. Indeed, imposing 
a simple MEM cut improves a traditional $m_{\ell\ell\gamma}$ fit by around a factor of two. The MEM@NLO 
was found to perform around 16\% better than the equivalent MEM@LO algorithm. Since 
this channel will require thousands of inverse femtobarns of data, such an improvement 
represents a significant benefit to the analysis. More recently the MEM has been used at LO 
to investigate the potential of finding off-shell Higgs events as a means
to constrain the Higgs width at the LHC~\cite{Campbell:2013una}. In this case the use of a MEM discriminant could improve 
limits obtained using a simple cut-and-count approach by a factor of $1.5$ or more. Given the importance of the Higgs width, and 
the inherent difficulty in measuring this in a hadronic environment, this again demonstrates the power 
of the MEM. 

\section{Conclusions}
\label{sec:Conc}

We have presented a method which reweights Born phase space points to next-to-leading order. 
The method relies on the generation of real phase configurations that are obtained by simple maps from an underlying Born 
topology. Each initial or final state parton branches, and the full real radiation phase space is recovered by the integration 
over all branchings. If this branching is inclusive then one naturally recovers the inclusive cross section.  Alternatively, 
if the emission is curtailed at some scale, one reproduces the exclusive NLO cross section.
Using our setup one can reproduce traditional NLO cross sections and distributions, by keeping 
all of the kinematic information regarding the branched final state. On the other hand, if one integrates out the 
branching, one can effectively define a dynamical $K$ factor for each event in the Born phase space. Since 
this integration requires mapping real emissions back to a LO topology, it is most natural to impose a scale at 
which the emission becomes too hard and is associated with an observed jet. By integrating out the emissions 
one obtains a procedure for weighting individual Born events in a way that accounts for
corrections from higher orders in perturbation theory.

\end{document}